\documentclass[twocolumn,aps,prc,showpacs,floatfix]{revtex4}
\usepackage{amsmath,bm}
\usepackage{graphicx}

\begin{document}

\title{Parton coalescence at RHIC}
\author{V. Greco}
\affiliation{Cyclotron Institute and Physics Department, Texas A\&M 
University, College Station, Texas 77843-3366, USA}
\author{C. M. Ko}
\affiliation{Cyclotron Institute and Physics Department, Texas A\&M 
University, College Station, Texas 77843-3366, USA}
\author{P. L\'evai}
\affiliation{KFKI Research Institute for Particle and Nuclear Physics,
P.O. Box 49, Budapest 1525, Hungary}

\date{\today}

\begin{abstract}
Using a covariant coalescence model, we study hadron production in 
relativistic heavy ion collisions from both soft partons in the quark-gluon 
plasma and hard partons in minijets. Including transverse flow of soft 
partons and independent fragmentation of minijet partons, the model is 
able to describe available experimental data on pion, kaon, and antiproton 
spectra. The resulting antiproton to pion ratio is seen to increase
at low transverse momenta and reaches a value of about one at 
intermediate transverse momenta, as observed in experimental data at 
RHIC. A similar dependence of the antikaon to pion ratio on transverse
momentum is obtained, but it reaches a smaller value at intermediate 
transverse momenta. At high transverse momenta, the model predicts  
that both the antiproton to pion and the antikaon to pion ratio 
decrease and approach those given by the perturbative QCD. Both 
collective flow effect and coalescence of minijet partons with partons in 
the quark-gluon plasma affect significantly the spectra of hadrons with 
intermediate transverse momenta. Elliptic flows of protons, Lambdas, and 
Omegas have also been evaluated from partons with elliptic flows 
extracted from fitting measured pion and kaon elliptic flows, and
they are found to be consistent with available experimental data.
\end{abstract}

\maketitle

\section{introduction}
Recently, there is a renewed interest in using the parton coalescence or 
recombination model to understand hadron production from the quark-gluon 
plasma formed in relativistic heavy ion collisions 
\cite{melting,lin,voloshin,hwa,fries,greco}. Emphases in these studies
are, however, different from earlies ones based on the coalescence 
model such as the ALCOR \cite{alcor} and the MICOR \cite{micro} model. 
Instead of addressing particle yields and their ratios, these new studies 
were more concerned with observables related to collective dynamics
and production of hadrons with relatively large transverse momentum.
In Ref. \cite{melting}, the parton coalescence was used to convert the 
quark matter, that is formed from melted soft strings produced in initial 
soft collisions, to hadrons. Including parton coalescence in a multiphase 
transport model (AMPT) \cite{ampt}, it was found that partonic effects are 
important for describing measured large elliptic flows and narrow two-pion 
correlation functions at RHIC.  In Ref. \cite{voloshin}, it was shown that 
hadron production based on parton coalescence is able to account for the 
qualitative difference between the observed elliptic flows of mesons and 
baryons. Based on parton recombination, a parton transverse momentum 
distribution is obtained in Ref. \cite{hwa} from the measured pion spectrum 
and is then used to predict the kaon and proton transverse momentum spectra. 
The parton coalescence model is further found in Refs. \cite{fries,greco} 
to be able to explain the observed enhancement of intermediate transverse 
momentum protons and antiproton. While coalescence of partons from a 
quark-gluon plasma with high effective temperature is considered in Ref. 
\cite{fries}, their coalescence with minijet partons is introduced in Ref. 
\cite{greco} as a new mechanism for hadronization of minijet partons. 

Besides independent fragmentations to hadrons as usually considered,
minijet partons produced in relativistic heavy ion collisions \cite{wang} 
are allowed in Ref. \cite{greco} to coalesce with partons from the 
quark-gluon plasma formed in relativistic heavy ion collisions, as 
suggested in Ref. \cite{lin} for studying the flavor ordering 
of the elliptic flows of hadrons with intermediate transverse momenta.
Since minijet partons have a power-law transverse momentum spectrum while 
partons in the quark-gluon plasma have an exponential thermal spectrum, 
this mechanism leads to an enhanced production of hadrons with intermediate 
transverse momentum. Because of stronger enhancement for baryons and 
antibaryons than for pions from this hadronization mechanism, a large 
antiproton to pion ratio of about one is obtained at intermediate 
transverse momenta as seen in the experimental data from the PHENIX 
collaboration \cite{ppi}. However, in this study hadrons from 
hadronization of the quark-gluon plasma are taken to also have 
exponential thermal spectra extending to all transverse momenta. Furthermore, 
in order to obtain a semi-analytical expression for the coalescence formula, 
only comoving partons at zero rapidity are considered in evaluating the 
coalescence probability for hadron production. Although relativistic 
kinematics was used in this study, the model is not fully covariant.
In the present paper, we relax these simplifications by using a covariant 
coalescence model and treating more generally parton coalescence via 
a Monte-Carlo method.  Also, we include coalescence among partons from the 
quark-gluon plasma so that hadrons with low momenta are treated on the 
same footing as those with intermediate momenta. We show that this improved
coalescence model for partons from both the quark-gluon plasma and minijets 
is able to reproduce the experimental transverse momentum spectra of pions,
antikaons, and antiprotons measured at RHIC as well as the antiproton to 
pion ratio as functions of transverse momentum. We further predict the 
dependence of the $K^-/\pi^-$ ratio on transverse momentum. Fitting  
quark elliptic flows to measured pion and kaon elliptic flows, the predicted 
elliptic flows of protons and $\Lambda$ are found to agree with available 
experimental data. The $\Omega$ elliptic flow is also predicted and is 
smaller than that of $\Lambda$ as a result of smaller elliptic flow for 
light quarks than strange quarks.

The paper is organized as follows. In Section \ref{coalescence}, we 
describe the general formalism of a covariant coalescence model for 
mesons and baryons. How minijet and quark-gluon plasma partons are 
determined for Au+Au collisions at 200 AGeV are described in Section 
\ref{partons}. In Section \ref{monte}, the Monte-Carlo method used for 
treating coalescence among partons whose numbers vary by many orders of 
magnitude is presented. Results for the transverse momentum spectra of pions, 
antiprotons, and antikaons obtained from the coalescence model are given in 
Section \ref{results}. Ratios of the antiproton to pion and the antikaon
to pion transverse momentum spectra are also shown. We further study the 
effect of coalescence of minijet partons with those in the quark-gluon 
plasma and also the effect of collective flow on the spectra and ratios of 
produced hadrons. The elliptic flow of hadrons based on that of quarks
is also studied. Finally, we conclude in Section \ref{summary} with a 
summary of present work and an outlook about future developments and 
applications of the parton coalescence model.

\section{the coalescence model}\label{coalescence}

Using the covariant coalescence model of Dover {\it et al.} \cite{dover}, 
the number of mesons formed from the coalescence of quark and antiquarks 
can be written as 
\begin{eqnarray}
N_M&=&g_M\int p_1\cdot d\sigma_1 p_2\cdot d\sigma_2
\frac{d^3{\bf p}_1}{(2\pi)^3E_1}\frac{d^3{\bf p}_2}{(2\pi)^3E_2}\nonumber\\
&\times&f_q(x_1;p_1)f_{\bar q}(x_2;p_2)f_M(x_1,x_2;p_1,p_2).
\label{coal}
\end{eqnarray}
In the above, $d\sigma$ denotes an element of a space-like hypersurface;
and $g_M$ is the statistical factor for forming a colorless meson 
from spin 1/2 color quark and antiquark. For mesons considered here, 
i.e., $\pi$, $\rho$, $K$, and $K^*$, the statistical factors are 
$g_\pi=g_K=1/36$ and $g_\rho=g_{K^*}=1/12$. The functions $f_q(x,p)$ and 
$f_{\bar q}(x,p)$ are, respectively, covariant distribution functions of 
quarks and antiquarks in the phase space, and they are normalized to their 
numbers, i.e.,
\begin{equation}  
\int p\cdot d\sigma\frac{d^3{\bf p}}{(2\pi)^3E}f_{q,\bar q}(x,p)=N_{q,\bar q}.
\end{equation}

The function $f_M(x_1,x_2;p_1,p_2)$ in Eq.(\ref{coal}) is the probability 
for a quark and an antiquark to form a meson. It describes the dynamic 
process of converting a quark and an antiquark to a bound state meson in 
the presence of a partonic matter. It depends on the overlap of the quark 
and antiquark wave functions with the wave function of the meson as well 
as the interactions of emitted virtual partons, which are needed for 
balancing the energy and momentum, with the partonic matter. Neglecting
the off-shell effects and taking the wave functions of quark and antiquark
to be plane waves, the coalescence probability function is then simply
the covariant meson Wigner distribution function. In Ref. \cite{dover}, 
it is parametrized by Gaussian functions in $x_1-x_2$ and $p_1-p_2$.
Here, we take it to have a uniform distribution as in Ref. \cite{greco}, 
i.e.,  
\begin{eqnarray}
&&f_M(x_1,x_2;p_1,p_2)=\frac{9\pi}{2(\Delta_x\Delta_p)^3}
\Theta\left(\Delta_x^2-(x_1-x_2)^2\right)\nonumber\\
&&\times\Theta\left(\Delta_p^2-(p_1-p_2)^2+(m_1-m_2)^2\right),
\end{eqnarray}  
where $\Delta_x$ and $\Delta_p$ are the covariant spatial and momentum 
coalescence radii. The factors in front of theta functions are 
introduced to obtain the correct normalization for the meson Wigner
function in the nonrelativistic limit. Here we use $\hbar=c=1$.

For ultrarelativistic heavy ion collisions at RHIC, it is convenient
to introduce rapidities variables $y$ and $\eta$ in the momentum and the 
coordinate space, respectively, and they are defined by 
\begin{eqnarray}
y=\frac{1}{2}\ln\frac{E+p_z}{E-p_z},\qquad 
\eta=\frac{1}{2}\ln\frac{t+z}{t-z}.
\end{eqnarray}
Using these variables, the spatial coordinate becomes 
$x=(\tau\cosh\eta,-\tau\sinh\eta,-{\bf r}_{\rm T})$, with 
$\tau=\sqrt{t^2-z^2}$ and ${\bf r}_{\rm T}$ denoting, respectively, 
proper time and transverse coordinates; while the momentum is 
$p=(m_{\rm T}\cosh y,-m_{\rm T}\sinh y,-{\rm p}_{\rm T})$ with
$m_{\rm T}=\sqrt{m_q^2+{\bf p}_{\rm T}^2}$ being the transverse mass 
in terms of quark mass $m_q$ and its transverse momentum ${\bf p}_{\rm T}$. 
The momentum volume element is then given by 
\begin{eqnarray}
\frac{d^3{\bf p}}{E}=dyd^2{\bf p}_{\rm T},
\end{eqnarray}
while the spatial volume element becomes
\begin{eqnarray}
p\cdot d\sigma=\tau m_{\rm T}\cosh(y-\eta)d\eta d^2{\bf r}_{\rm T},
\end{eqnarray}
if we adopt a hypersurface of constant longitudinal proper times.

The quark and antiquark phase space distribution functions are then
given by 
\begin{equation}
f_{q,\bar q}(x,p)=\frac{(2\pi)^3}{\tau m_{\rm T}\cosh(y-\eta)}
\frac{dN_{q,\bar q}}{d\eta d^2{\bf r}_{\rm T}dyd^2{\bf p}_{\rm T}},
\end{equation}
where $R_\perp$ is the transverse radius of the system.

The yield of mesons from coalescence of quarks and antiquarks is then 
given by 
\begin{eqnarray}
N_M&=&g_M\int d\eta_1d^2{\bf r}_{1{\rm T}}d\eta_2d^2{\bf r}_{2{\rm T}} 
dy_1d^2{\bf p}_{1{\rm T}}dy_2d^2{\bf p}_{2{\rm T}}\nonumber\\
&\times&\frac{dN_q}{d\eta_1d^2{\bf r}_{1{\rm T}}dy_1d^2{\bf p}_{1{\rm T}}}
\frac{dN_{\bar q}}{d\eta_2d^2{\bf r}_{2{\rm T}}dy_2d^2{\bf p}_{2{\rm T}}}
\nonumber\\
&\times&f_M(x_1,x_2;p_1,p_2).
\end{eqnarray}

The transverse momentum spectrum of mesons can be obtained from 
Eq.(\ref{meson}) by multiplying the right hand side with
\begin{eqnarray}
1=\int d^2{\bf p}_{\rm T}\delta^{(2)}({\bf p}_{\rm T}-{\bf p}_{1{\rm T}}
-{\bf p}_{2{\rm T}}),
\end{eqnarray} 
and then differentiating both sides of the equation with respect to
${\bf p}_{\rm T}$. The resulting meson transverse momentum spectrum from 
quark and antiquark coalescence is given by  
\begin{eqnarray}
&&\frac{dN_M}{d^2{\bf p}_{\rm T}}=g_M\int d\eta_1d^2{\bf r}_{1{\rm T}}
d\eta_2d^2{\bf r}_{2{\rm T}}dy_1d^2{\bf p}_{1{\rm T}}\nonumber\\
&&\times dy_2d^2{\bf p}_{2{\rm T}}
\frac{dN_q}{d\eta_1d^2{\bf r}_{1{\rm T}}dy_1d^2{\bf p}_{1{\rm T}}}
\frac{dN_{\bar q}}{d\eta_2d^2{\bf r}_{2{\rm T}}dy_2d^2{\bf p}_{2{\rm T}}}
\nonumber\\
&&\times f_M(x_1,x_2;p_1,p_2)
\delta^{(2)}({\bf p}_{\rm T}-{\bf p}_{1{\rm T}}-{\bf p}_{2{\rm T}}).
\end{eqnarray}

For quarks and antiquarks produced at central rapidities in relativistic heavy 
ion collisions, it is reasonable to assume that their longitudinal momentum
distributions are boost-invariant, i.e., independent of rapidity. Furthermore, 
they are expected to satisfy the Bjorken correlation of equal spatial and 
momentum rapidities, i.e., $\eta=y$. The quark and antiquark phase space 
distribution functions in the rapidity range $\Delta y$ can then be 
expressed as
\begin{eqnarray}
\frac{dN_{q,\bar q}}{d\eta d^2{\bf r}_{\rm T}dyd^2{\bf p}_{\rm T}}=
\frac{\delta(\eta-y)}{\Delta y}
\frac{dN_{q,\bar q}}{d^2{\bf r}_{\rm T}d^2{\bf p}_{\rm T}}
\Bigl|_{|y|\le\Delta y/2}.
\end{eqnarray}
This leads to the following meson transverse momentum spectrum from 
coalescence of quarks and antiquarks
\begin{eqnarray}\label{meson}
\frac{dN_M}{d^2{\bf p}_{\rm T}}&=&\frac{g_M}{(\Delta y)^2}
\int d^2{\bf r}_{1{\rm T}}d^2{\bf r}_{2{\rm T}}
d^2{\bf p}_{1{\rm T}}d^2{\bf p}_{2{\rm T}}\nonumber\\
&\times&\frac{dN_q}{d^2{\bf r}_{1\rm T}d^2{\bf p}_{1\rm T}}
\Bigl|_{|y_1|\le\Delta y/2} 
\frac{dN_{\bar q}}{d^2{\bf r}_{2\rm T}d^2{\bf p}_{2\rm T}}
\Bigl|_{|y_2|\le\Delta y/2}\nonumber\\
&\times&\int d\eta_1dy_d\eta_2dy_2\delta(\eta_1-y_1)\delta(\eta_2-y_2)
\nonumber\\
&\times&f_M(x_1,x_2;p_1,p_2)
\delta^{(2)}({\bf p}_{\rm T}-{\bf p}_{1{\rm T}}-{\bf p}_{2{\rm T}}).
\end{eqnarray}

The above result can be simplified if there is no correlation between 
parton transverse momenta and positions, such as in the absence of collective 
transverse flow, and if partons are also uniformly distributed in the 
transverse space. In this case, the quark and antiquark distributions only 
depend on transverse momentum, i.e., 
\begin{eqnarray}
\frac{dN_{q,\bar q}}{d^2{\bf r}_{\rm T}d^2{\bf p}_{\rm T}}
\Bigl|_{|y|\le\Delta y/2}
=\frac{1}{\pi R_\perp^2}\frac{dN_{q,\bar q}}{d^2{\bf p}_{\rm T}}
\Bigl|_{|y|\le\Delta y/2}.
\end{eqnarray}
Considering small rapidity range such as $\Delta y\le 1$, we have 
\begin{eqnarray}
(x_1-x_2)^2&=&2\tau^2[1-\cosh(\eta_1-\eta_2)]-({\bf r}_{1{\rm T}}
-{\bf r}_{2{\rm T}})^2\nonumber\\
&\approx&-\tau^2(\eta_1-\eta_2)^2
-({\bf r}_{1{\rm T}}-{\bf r}_{2{\rm T}})^2\nonumber\\
&\approx& -({\bf r}_1-{\bf r}_2)^2,
\end{eqnarray}
and
\begin{eqnarray}
(p_1-p_2)^2&=&m_{1{\rm T}}^2+m_{2{\rm T}}^2-2m_{1{\rm T}}m_{2{\rm T}}
\cosh(y_1-y_2)\nonumber\\
&-&({\bf p}_{1{\rm T}}-{\bf p}_{2{\rm T}})^2\nonumber\\
&\approx&(m_{1{\rm T}}-m_{2{\rm T}})^2-({\bf p}_{1{\rm T}}
-{\bf p}_{2{\rm T}})^2. 
\end{eqnarray}
After carrying out the integrals in transverse space, the meson transverse 
momentum spectra is then give by 
\begin{eqnarray}\label{prl}
&&\frac{dN_M}{d^2{\bf p}_{\rm T}}=g_M\frac{6\pi}{\tau\Delta y
R_\perp^2\Delta_p^3}\int d^2{\bf p}_{1{\rm T}}d^2{\bf p}_{2{\rm T}}
\nonumber\\
&&\times\frac{dN_q}{d^2{\bf p}_{1{\rm T}}}\Bigl|_{|y_1|\le\Delta y/2}
\frac{dN_{\bar q}}{d^2{\bf p}_{2{\rm T}}}\Bigl|_{|y_2|\le\Delta y/2}\nonumber\\
&&\times\delta^{(2)}({\bf p}_{\rm T}-{\bf p}_{1{\rm T}}-{\bf p}_{2{\rm T}})
\Theta\left(\Delta_p^2-({\bf p}_{1{\rm T}}-{\bf p}_{2{\rm T}})^2
\right.\nonumber\\
&&\left.-[(m_{1{\rm T}}-m_{2{\rm T}})^2-(m_1-m_2)^2]\right).
\end{eqnarray}
This result for meson transverse momentum spectra reduces to that used in 
our previous schematic study \cite{greco} if we assume that only comoving 
quarks and antiquarks can coalesce to hadrons and replace the arguments 
of theta functions in Eq.(\ref{prl}) by $\Delta_p^2$ minus quark and 
antiquark relative momentum in meson rest frame. 

Since we will include collective transverse flow of partons in the
quark-gluon plasma, Eq.(\ref{meson}) will be used in the following 
study. To generalize the results for mesons to formation of baryons and 
antibaryons from the parton distribution functions, we take the baryon 
coalescence probability function as
\begin{eqnarray}
&&F_B(x_1,x_2,x_3;p_1,p_2,p_3)\nonumber\\
&&=\frac{9\pi}{2\Delta_x^3\Delta_p^3}
\Theta\left(\Delta_x^2-\frac{1}{2}(x_1-x_2)^2\right)\nonumber\\
&&\times\Theta\left(\Delta_p^2-\frac{1}{2}(p_1-p_2)^2\right)\nonumber\\
&&\times\frac{9\pi}{2\Delta_x^3\Delta_p^3}
\Theta\left(\Delta_x^2-\frac{1}{6}(x_1+x_2-2x_3)^2\right)\nonumber\\
&&\times\Theta\left(\Delta_p^2-\frac{1}{6}[(p_1+p_2-2p_3)^2\right.\nonumber\\
&&\left.-(m_1+m_2-2m_3)^2]\right),
\end{eqnarray} 
where we have taken for simplicity the same space and momentum coalescence 
radii for the relative Jacobi coordinates among three quarks. 

For boost-invariant dynamics with Bjorken spatial and momentum rapidities 
correlation, we obtain following baryon transverse momentum spectrum 
from coalescence of three quarks:
\begin{eqnarray}\label{baryon}
&&\frac{dN_B}{d^2{\bf p}_{\rm T}}=\frac{g_B}{(\Delta y)^3}
\int\prod_{i=1}^3d^2{\bf r}_{i\rm T}d^2{\bf p}_{i{\rm T}}
\frac{dN_q}{d^2{\bf r}_{i\rm T}d^2{\bf p}_{i{\rm T}}}
\Bigl|_{|y_i|\le\Delta y/2}\nonumber\\
&&\times\int\prod_{i=1}^3d\eta_idy_i\delta(\eta_i-y_i)
F_B(x_1,x_2,x_3;p_1,p_2,p_3)\nonumber\\
&&\times\delta^{(2)}\left({\bf p}_{\rm T}
-\sum_{i=1}^3{\bf p}_{i{\rm T}}\right).
\end{eqnarray}
In the above, $g_B$ is the statistical factor for formation of
a baryon from three quarks. For baryons and antibaryons considered 
in present study, i.e., $p$, $\Delta$, $\bar p$ and $\bar\Delta$, the 
statistical factors are $g_p=g_{\bar p}=1/108$ and 
$g_\Delta=g_{\bar\Delta}=1/54$. The above formula can also be used 
for antibaryons by replacing quark momentum spectra by the momentum 
spectra of antiquarks.

\section{parton distributions}\label{partons}

We consider central Au+Au collisions ($0-10\%$) at 200 AGeV, available
at RHIC. In these collisions, hard processes between initial nucleons 
lead to production of minijets with large transverse momentum. Also,
a quark-gluon plasma is expected to be formed from soft processes
in the collisions. In this section, both parton momentum spectra 
in minijets and in the quark-gluon plasma are introduced.

\subsection{minijet partons}

Partons at high transverse momenta (usually greater than 2 GeV) are
mainly from the minijets produced in initial hard collisions among 
nucleons. The transverse momentum distribution of minijet partons in the 
midrapidity can be obtained from an improved perturbative QCD calculation 
\cite{yi02}. It is given by $dN_{\rm jet}/d^2p_T=1/\sigma^{0-10}_{\rm tot} 
d\sigma_{\rm jet}/d^2p_T$ in terms of $\sigma^{0-10}_{\rm tot}$ corresponding
to the total cross section at central 10\% of the collisions and the
jet production cross section from nucleus-nucleus collisions,
\begin{eqnarray}
\frac{d\sigma_{\rm jet}}{d^2{\bf p}_{\rm T}} \ &=& \ 
\int d^2{\bf b}\ d^2{\bf r}\ t_{\rm Au}({\bf r}) t_{\rm Au}(\bf{b}-\bf{r}) 
\nonumber \\ 
&\times&\sum_{\!\!ab}\!\!\int\!\!dx_a dx_b d^2{\bf k_a}_{\rm T} 
d^2{\bf k_b}_{\rm T} g({\bf k_a}_{\rm T}) g({\bf k_b}_{\rm T})\nonumber \\
&\times&f_{\!a/{\rm Au}}(x_a,Q^2)f_{\!b/{\rm Au}}(x_b,Q^2)\nonumber\\
&\times&\frac{\hat s}{\pi}\delta(\hat s+\hat t+\hat u)
\frac{d\sigma^{ab}}{d\hat t}.
\end{eqnarray}
In the above, $t_{\rm A}({\bf r})$ is the thickness function of Au at 
transverse radius ${\bf r}$ given by integrating  the nuclear density
distribution along the longitudinal direction.  The parton distribution 
function in a nucleon in the nucleus Au is denoted by 
$f_{a/{\rm Au}}(x,Q^2)$ including transverse smearing $g({\bf k}_{\rm T})$. 
The cross section $d\sigma^{ab}/d\hat t$ is the parton scattering
cross section. Kinematic details and a systematic analysis of $pp$
collisions can be found in Ref.~\cite{yi02}. Using the GRV94 LO result 
for the PDF \cite{structure} and the KKP fragmentation function from 
Ref. \cite{KKP}, measured data in the reaction $pp\to\pi^0X$ at 
$\sqrt{s}=200$ GeV can be reproduced with $Q=0.75 p_{\rm T}$ and 
$\langle k_{\rm T}^2 \rangle = 2$ GeV$^{2}$. 

In heavy ion collisions at RHIC, minijet partons are expected to lose 
energy by radiating soft partons as they traverse through the quark-gluon 
plasma. This effect can be taken into account by lowering their transverse
momenta by the energy loss $\Delta E$, which depends on both the parton
energy $E$ and an effective opacity parameter $L/\lambda$ according to 
the GLV model \cite{levai1}.  

\begin{figure}[ht]
\includegraphics[height=3in,width=2.8in,angle=-90]{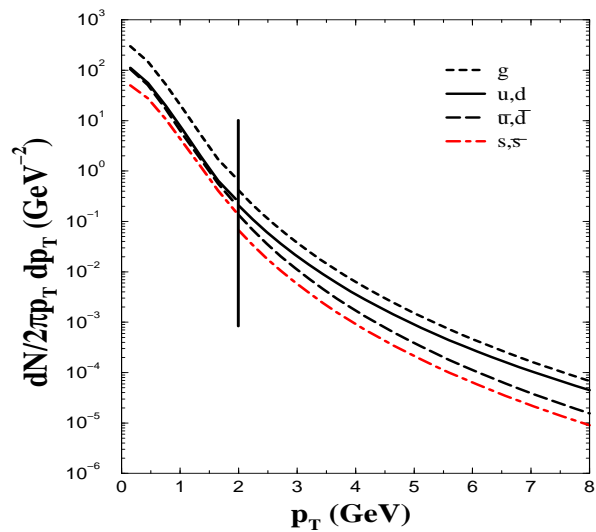}
\vspace{-0.2cm}
\caption{Transverse momentum distributions of partons at hadronization in 
Au+Au collisions at $\sqrt s=200$ AGeV for gluons (short-dashed curve), 
$u$ and $d$ (solid curve), $\bar u$ and $\bar d$ (dashed curve) 
as well as $s$ and $\bar s$ (dash-dotted curve) quarks.
Minijet partons have transverse momenta greater than 2 GeV, while partons 
from the quark-gluon plasma have transverse momenta below 2 GeV.}
\label{parton}
\end{figure}

Taking as momentum cutoff $p_0=2$ GeV/c for minijet partons and using an 
effective opacity $L/\lambda=3.5$ as extracted from fitting the spectrum 
of high transverse momentum pions measured at RHIC \cite{levai1,gyulassy1}, 
the transverse momentum spectra of minijet partons at midrapidity ($y=0$) 
in central Au+Au collisions at $\sqrt s=200$ AGeV are shown in 
Fig. \ref{parton} for gluons (short-dashed curve), $u$ and $d$  
(solid curve), $\bar u$ and $\bar d$ (dashed curve) as well as
$s$ and $\bar s$ (dash-dotted curve) quarks. These spectra can 
be parametrized as
\begin{equation}\label{parton-q}
\frac{dN_{\rm jet}}{d^2{\bf p}_{\rm T}}=
A\left(\frac{B}{B+p_{\rm T}}\right)^n,
\end{equation}
Values for the parameters $A$, $B$, and $n$ for gluons, light and strange 
quarks and anitquarks are given in Table \ref{tab:minijets}. For later 
calculation of parton coalescence probability, masses of minijet partons 
are taken to be the current quark masses, i.e., 10 MeV for light quarks 
and 175 MeV for strange quarks. Also, we assume that the rapidity 
distribution of minijet partons is uniform in the rapidity range of 
$y\in(-0.5,+0.5)$ considered in the present study. 

\begin{table}[th]
\caption{Parameters for minijet parton distributions given in 
Eq.(\ref{parton-q}) at midrapidity from  Au+Au at $\sqrt{s}=200$ GeV.}
\medskip
\begin{tabular}{cccc}\hline\hline
\hfil & $\quad A[1/{\rm GeV}^2]\quad$ & $\quad B[{\rm GeV}]\quad$ & 
$\quad n\quad$ \\ \hline
    g & 3.2$\times 10^4$ & 0.5 & 7.1 \\ \hline
  u,d & 9.8$\times 10^3$ & 0.5 & 6.8 \\ \hline
$\bar u,\bar d$ & 1.9$\times 10^4$ & 0.5 & 7.5 \\ \hline
s,$\bar s$ & 6.5$\times 10^3$ & 0.5 & 7.4 \\ \hline\hline
\end{tabular}
\label{tab:minijets}
\end{table}

\subsection{the quark-gluon plasma}

For partons in the quark-gluon plasma, their transverse momentum
spectra are taken to have an exponential form. For their longitudinal
momentum distribution, we assume that they are boost-invariant, i.e.,
they have a uniform rapidity distribution in the range $y\in(-0.5,+0.5)$.
To take into account collective flow of quark-gluon plasma, these
partons are boosted by a flow velocity 
${\bf v}_{\rm T}=\beta_0({\bf r}_{\rm T}/R_\perp)$, 
depending on their transverse radial positions $r_{\rm T}$. Here, 
$R_\perp$ is the transverse size of the quark-gluon plasma at hadronization, 
and $\beta_0$ is the collective flow velocity of the quark-gluon plasma and 
is taken to be 0.5c. In this case, the light quarks and antiquarks transverse
momentum spectra are given by 
\begin{eqnarray}\label{quark}
\frac{dN_{\rm q,\bar q}}{d^2{\bf r}_{\rm T}d^2{\bf p}_{\rm T}}
&=&\frac{g_{q,\bar q}\tau m_{\rm T}}{(2\pi)^3}\nonumber\\
&\times&\exp\left(-\frac{\gamma_{\rm T}(m_{\rm T}-{\bf p}_{\rm T}\cdot
{\bf v}_{\rm T}\mp\mu_q)}{T}\right),
\end{eqnarray}
where $g_q=g_{\bar q}=6$ are the spin-color degeneracy of light quarks
and antiquarks, and the minus and plus signs are for quarks and antiquarks, 
respectively. The slope parameter $T$ is taken to be $T=170$ MeV, 
consistent with the phase transition temperature ($T\sim 165-185$ MeV) 
from lattice QCD calculations. Masses of light quarks and antiquarks are 
taken to be $m_q=m_{\bar q}=300$ MeV, similar to the masses of constituent 
quarks due to possible nonperturbative effects in the quark-gluon plasma near
hadronization \cite{levai}. For the quark chemical potential $\mu_q$, we 
use a value of $\mu_q=10$ MeV to give a light antiquark to quark ratio of 
0.89, which would then lead to an antiproton to proton ratio of about 
$(0.89)^3=0.7$, consistent with the observed ratio at midrapidity in heavy 
ion collisions at RHIC. The transverse flow effect is taken into account
through the factor $\gamma_{\rm T}=1/\sqrt{1-v_{\rm T}^2}$.

The momentum spectra for strange quarks and antiquarks are similar 
to Eq.(\ref{quark}) with $m_q$ replaced by the strange quark mass 
$m_s=m_{\bar s}=475$ MeV and $\mu_q$ replaced by $\mu_q-\mu_s$.
The strange chemical potential $\mu_s$ is taken to have the same value
as $\mu_q$ in order to have same strange and antistrange quarks numbers.
The resulting strange quark to light quark ratio is then about 0.27.
Including contribution from decays of $\rho$ and $K^*$, this would then 
give a final $K^-/K^+=0.89$ and $K^-/\pi^-=0.24$, comparable to experimental
data at midrapidity from RHIC. Eq. (\ref{quark}) also applies to gluons 
after replacing $g_q$ by the gluon spin-color degeneracy $g_g=16$ and 
dropping the chemical potential. For the gluon mass, we take it
to be similar to that of light quarks in order to take into account
nonperturbative effects in the quark-gluon plasma.

Although the shape of quark-gluon plasma is nearly cylindrically 
symmetric, its parton momentum distribution is azimuthally asymmetric 
as shown by the finite elliptic flow of hadrons observed in experiments. 
Since we are interested in hadron spectra that are integrated over the 
azimuthal angle, including such asymmetry is not expected to change much 
our results. Nonetheless, the Monte-Carlo method to be described in 
Section \ref{monte} to evaluate the coalescence formula can be easily 
extended to take into account such effects in order to study the relation 
between the elliptic flow of partons and those of hadrons as suggested in 
Ref. \cite{voloshin} and to be studied in Section \ref{elliptic}.

The quark-gluon plasma is further assumed to have a transverse 
radius of $R_\perp=8.3$ fm at proper time $\tau=4$ fm, corresponding to 
a volume of $V=900$ fm$^3$. Positions of partons in the transverse
direction are taken to have a uniform distribution. Their longitudinal 
positions are then determined by $z=\tau\sinh y$, as we have assumed 
that $\eta=y$ due to assumed Bjorken correlation. The resulting total 
transverse energy per unit rapidity from both the expanding quark-gluon 
plasma and minijet partons is about 590 GeV and is consistent 
with that measured by the PHENIX collaboration \cite{etr}. Most of this 
transverse energy comes from soft QGP partons as the contribution of 
minijet partons is only about 10\%. The resulting parton density is about 
$\rho_{\rm parton}\sim 1$ fm$^{-3}$.  The thermal parton spectra 
below the momentum cutoff $p_0=2$ GeV from the quark-gluon plasma including 
the collective flow effect are shown in Fig. \ref{parton} for gluons 
(short-dashed curve), $u$ and $d$ (solid curve), $\bar u$ 
and $\bar d$ (dashed curve) as well as $s$ and $\bar s$ 
(dash-dotted curve) quarks. Because of scattering of minijet partons with 
thermal partons as they traverse the quark-gluon plasma, those with momentum 
around $p_0$ are expected to be thermalized with partons in the quark-gluon 
plasma, leading to a smooth spectrum around $p_0$. In the present study, 
we neglect this effect. 

\section{the Monte-Carlo method}\label{monte}

In our previous study \cite{greco}, only partons at midrapidity ($y=0$) 
are considered. Furthermore, only comoving partons, i.e., partons with 
momenta in the same transverse direction are allowed to coalesce to 
hadrons. In these limits, the coalescence formula is reduced to a
one-dimensional integral for mesons and a two-dimensional integral 
for baryons. In the present study, we do not introduce these simplifications. 
Instead, the multi-dimensional integrals in the coalescence formula,
given by Eqs.(\ref{meson}) and (\ref{baryon}) are evaluated by the 
Monte-Carlo method via test particles. Specifically, we introduce a large 
number of test partons with uniform momentum distribution. To take into 
account the large difference between numbers of thermal and minijet 
partons, a test parton with momentum ${\bf p}_{\rm T}$ is given a 
probability that is proportional to the parton momentum distribution, 
e.g., ${dN_q}/{d^2{\bf p}_{\rm T}}$ for quarks, with the proportional 
constant determined by requiring that the sum of all parton probabilities 
is equal to the parton number. With test partons, the coalescence formulas, 
Eqs.(\ref{meson}) and (\ref{baryon}), are rewritten as
\begin{eqnarray}
\frac{dN_M}{d^2{\bf p}_{\rm T}}&=&g_M
\sum_{i,j}P_q(i)P_{\bar q}(j)\delta^{(2)}({\bf p}_{\rm T}-{\bf p}_{i{\rm T}}
-{\bf p}_{j{\rm T}})\nonumber\\
&\times&f_M(x_i,x_j;p_i,p_j).
\end{eqnarray}
and
\begin{eqnarray}
\frac{dN_B}{d^2{\bf p}_{\rm T}}&=&g_B
\sum_{i\ne j\ne k}P_q(i)P_q(j)P_q(k)\nonumber\\
&\times&\delta^{(2)}({\bf p}_{\rm T}-{\bf p}_{i{\rm T}}-{\bf p}_{j{\rm T}}-
{\bf p}_{k{\rm T}})\nonumber\\
&\times&f_B(x_i,x_j,x_k;p_i,p_j,p_k).
\end{eqnarray}
In the above, $P_q(i)$ and $P_{\bar q}(j)$ are probabilities carried by 
$i$th test quark and $j$th test antiquark. 

The Monte-Carlo method introduced here allows us to treat the coalescence 
of low momentum partons on the same footing as that of high momentum ones. 
We find that despite a decrease of eight orders-of-magnitude in real 
particle spectra, about equal numbers of test hadrons are formed at all 
momenta.

\section{hadron transverse momentum spectra}\label{results}

In this section, we show results for the transverse momentum spectra
of pions, antiprotons, and antikaons using the model described in 
previous sections. For the coalescence contribution, we first 
take into account the effects due to gluons in the quark-gluon plasma 
and minijets by converting them to quarks and antiquark pairs 
with probabilities according to the flavor compositions in
the quark-gluon plasma, as assumed in the ALCOR model \cite{alcor}.  
For both mesons and baryons, we include not only coalescence of hard 
and soft partons as in Ref.\cite{greco} but also that among soft 
as well as hard partons. Furthermore, we include
stable hadrons such as pion, nucleon (antinucleon), and kaon (antikaon) 
as well as unstable resonances such as $\rho$, $\Delta$, and $K^*$. 
Since the coalescence model can be viewed as formation of bound states
from interacting particles with energy mismatch balanced by other 
particles in the system, neglecting such off-shell effects is reasonable 
if the energy mismatch is small. The model is thus applicable for rho and 
nucleon (antinucleon) production when we take into account quark
masses. For other hadrons, the coalescence probability is expected to 
be reduced as a result of energy mismatch. The reduction factor can 
in principle be estimated by evaluating the transition probability in 
the presence of other particles. Since the coalescence radii $\Delta_x$ 
and $\Delta_p$ are treated as parameters in our study, the off-shell 
effects can be phenomenologically taken into account by using different 
coalescence radii for different hadrons. However, for simplicity we use same 
spatial coalescence radius $\Delta_x=0.85$ fm and momentum coalescence 
radius  $\Delta_p=0.24$ for both mesons and baryons. We note that there 
are more $\rho$ and $K^*$ than pion and kaon produced from parton 
coalescence. The pion, proton, and kaon transverse momentum spectra from 
parton coalescence shown below include contributions from decays of rho, 
$K^*$, and $\Delta$.

We have also included contributions to hadron production from  
minijet fragmentations. These are obtained using the KKP fragmentation 
function \cite{KKP}, which has been shown to reproduce measured high 
transverse momentum particles at RHIC. Explicitly, hadron momentum
spectra are related to minijet parton momentum spectra by 
\begin{eqnarray}
\frac{dN}{d^2{\bf p}_{\rm had}}=\sum_{\rm jet}\int dz \frac{dN}
{d^2{\bf p}_{\rm jet}}\frac{D_{\rm had/jet}(z,Q^2)}{z^2},
\end{eqnarray}
where $z=p_{\rm jet}/p_{\rm had}$ is the fraction of minijet momentum 
carried by the formed hadron and $Q=p_{\rm had}/2z$ is the momentum
scale for the fragmentation process. The KKP fragmentation function
is denoted by $D_{\rm had/jet}(z,Q^2)$.

Before we show the results for hadron spectra and elliptic flows,
we would like to point out that the coalescence model as formulated
here is applicable if the number of hadrons produced is much less
than the parton numbers. In this respect, results for hadrons with
momenta above about 1 GeV/c, which account for about 5\% of all
partons, are reliable. For lower momentum hadrons, one needs to 
impose conservation of parton number in converting them to 
hadrons via coalescence. This correction has not been included in
present study.

\subsection{pion transverse momentum spectrum}

\begin{figure}[ht]
\vspace{-0.2cm}
\includegraphics[height=3in,width=2.5in,angle=-90]{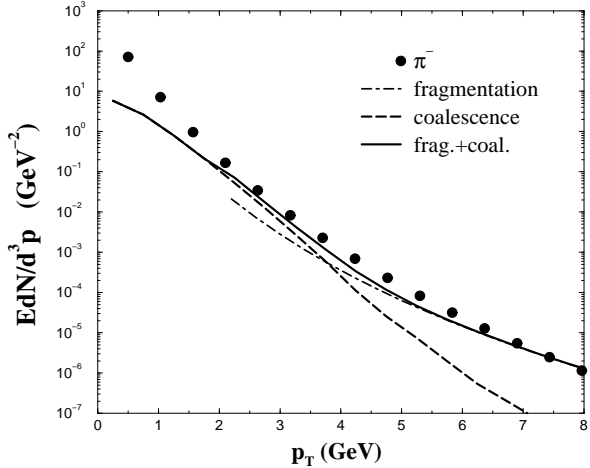}
\caption{Pion transverse momentum spectra from Au+Au collisions at
$\sqrt s=200$ AGeV: direct pion production from parton coalescence 
(dashed curve); pions from minijet fragmentations (dash-dotted curve);
and sum from the above two contributions (solid curve). Parametrized 
experimental $\pi^0$ data \cite{phenix1} are shown by filled circles.}
\label{pion-nd}
\end{figure}

In Fig. \ref{pion-nd}, we show the transverse momentum spectrum of pions
formed directly from parton coalescence (dashed curve). Pions from 
fragmentations of minijet partons are shown by dash-dotted curve. It 
is seen that pions from parton coalescence dominate low transverse 
momenta while those from minijet fragmentations are important at 
high transverse momenta. The two contributions have a similar magnitude 
at transverse momentum of about 3 GeV. Also shown in the figure is the 
total pion transverse momentum spectrum from the two contributions 
(solid curve). Compared with measured spectrum from the PHENIX 
Collaboration \cite{phenix1} (filled circles), the predicted spectrum
of directly produced pion at low transverse momenta is much below 
experimental data. 

\begin{figure}[ht]
\includegraphics[height=3in,width=2.5in,angle=-90]{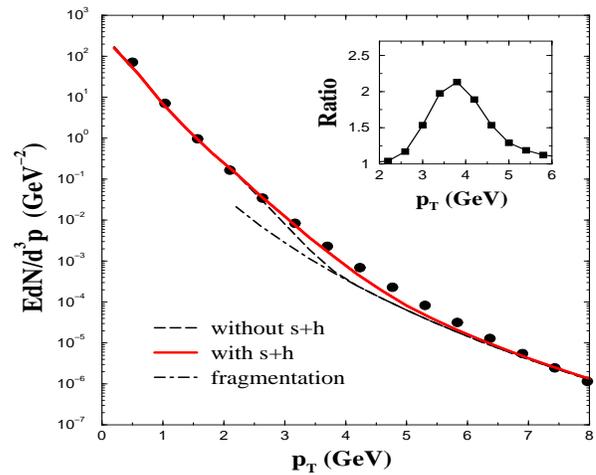}
\vspace{-0.2cm}
\caption{Pion transverse momentum spectra from Au+Au collisions at 
$\sqrt s=200$ AGeV with (solid curve) and without (dashed curve)
contributions from coalescence of minijet partons with the quark-gluon 
plasma partons. Parametrized experimental $\pi^0$ data \cite{phenix1} are 
shown by filled circles. Ratio of the two spectra is given in the inset.}
\label{pion-t}
\end{figure}

Since $\rho$, $K^*$, and $\Delta$ decay to pions, we have also included 
their contributions to the pion transverse momentum spectrum. It is found
that rho meson decays contribute significantly to the pion spectrum at low 
transverse momenta and bring the final pion spectrum, shown by the 
solid curve in Fig. \ref{pion-t}, in good agreement with experimental 
data. Also shown in this figure by the dashed curve is the pion spectrum 
without including contribution from coalescence of minijet partons with
partons in the quark-gluon plasma. It underestimates the pion spectrum 
at intermediate transverse momenta around 3.5 GeV/c. The contribution 
from coalescence of partons from minijets with those from the quark-gluon 
plasma is more clearly seen from the ratio of pion spectra with and without 
this contribution, shown in the inset of Fig. \ref{pion-t}, which is
more than a factor of 2 at transverse momenta around 4 GeV/c. Our results 
thus confirm our previous results in Ref.\cite{greco} based on a 
parametrized pion spectrum from hadronization of the quark-gluon plasma.

\subsection{antiproton transverse momentum spectrum}

\begin{figure}[ht]
\includegraphics[height=3in,width=2.5in,angle=-90]{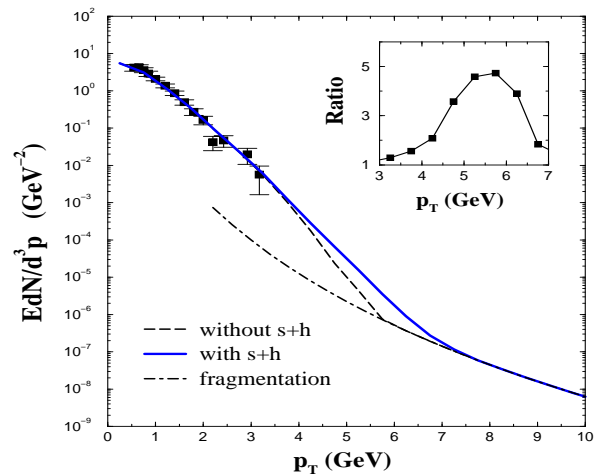}
\vspace{-0.2cm}
\caption{Transverse momentum spectra of antiprotons from Au+Au collisions
$\sqrt s=200$ AGeV. The dashed curve includes hadrons only from 
coalescence of partons in the quark-gluon plasma and from independent 
fragmentations of minijet partons (dash-dotted curve). Adding also hadrons 
from coalescence of minijet partons with partons in the quark-gluon plasma 
gives the solid curve. Ratio of the solid to the dashed curve is given in 
the inset. Experimental $p^-$ data \cite{phenix} at 130 AGeV are shown by 
filled circles.}
\label{aprot}
\end{figure}

In Fig. \ref{aprot}, we show the antiproton spectrum including those
from decays of $\bar\Delta$ for Au+Au collisions at $\sqrt s=200$ AGeV. 
Results for both with (solid curve) and without (dashed curve) contributions 
from coalescence of minijet partons with partons from the quark-gluon plasma
are shown. They include contributions from fragmentations of minijet 
partons (dash-dotted curve). The contribution from coalescence of
soft and hard partons becomes important when the transverse momentum 
is above 3 GeV/c, which is somewhat higher than in the case of pion
transverse momentum spectrum. Since there are no published experimental
data for antiproton transverse momentum spectrum from Au+Au collisions
at 200 AGeV, we compare our predictions with the experimental data 
from the PHENIX collaboration for Au+Au collisions at $\sqrt s=130$ AGeV 
\cite{phenix}, shown by filled squares for transverse momenta below 3 GeV/c.
Both predictions with and without contributions from coalescence of 
minijet partons with partons from the quark-gluon plasma are comparable
to the experimental data. Shown in the inset of this figure is their 
ratio as a function of transverse momentum. It is seen that the ratio 
is close to a factor of 5 at transverse momenta around 5.5 GeV/c. The 
contribution from coalescence of minijet partons with those from the 
quark-gluon plasma is thus more important for antiprotons than for pions. 
It would be of great interest to have experimental data for antiprotons 
at such high transverse momenta to verify our predictions. 

\subsection{antiproton to pion ratio}

\begin{figure}[ht]
\includegraphics[height=3in,width=2.5in,angle=-90]{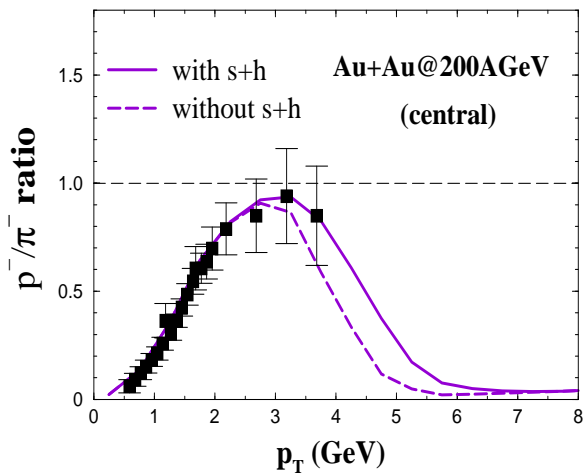}
\vspace{-0.2cm}
\caption{Antiproton to pion ratio from Au+Au collisions at $\sqrt s=200$ 
AGeV. Solid and dashed curves are, respectively, results with and
without contributions to antiprotons from coalescence of minijet partons 
with those from the quark-gluon plasma. Filled squares are the experimental 
data \cite{ppi}.}
\label{ratio}
\end{figure}

The antiproton to pion ratio is shown in Fig. \ref{ratio} as a function 
of transverse momentum. The solid curve is the result including 
contributions from both parton coalescence and minijet fragmentations.
The ratio increases with transverse momentum up to about 3 GeV/c and 
decreases with further increasing transverse momentum as in the experimental 
data \cite{ppi} shown by filled squares. At high transverse momenta, 
the antiproton to pion ratio becomes very small as it is largely determined 
by the results from pQCD \cite{zhang}. Neglecting the contribution to
aniproton production from coalescence of minijet partons with partons 
from the quark-gluon plasma reduces the antiproton to pion ratio at 
transverse momenta above 2.5 GeV as shown by the dashed curve in Fig. 
\ref{ratio}.

The enhanced antiproton to pion ratio at intermediate transverse momenta
was previously attributed to antiproton production from the baryon 
junctions in incident nucleons \cite{vitev}. The possibility of 
enhanced baryon to pion ratio due to parton coalescence was suggested 
in Ref. \cite{voloshin}. Using a parton distribution function that is 
fitted to measured pion transverse momentum spectrum, a parton 
recombination model similar to the coalescence model indeed leads 
to a large antiproton to pion ratio at intermediate transverse momenta
\cite{hwa}. In Ref. \cite{fries}, the antiproton to pion anomaly is
explained by the recombination of partons from a quark-gluon plasma
with a high effective temperature. Our model further introduces 
coalescence of minijet partons with partons in the quark-gluon 
plasma. This makes it possible to account for both the large 
antiproton to pion ratio at intermediate transverse momenta and its 
behavior at low transverse momenta.

In our previous work \cite{greco}, an increasing antiproton to pion ratio 
at low transverse momenta was also seen, but it was obtained by using 
different inverse slope parameters for antiproton and pion transverse 
momentum spectra. In contrast, results obtained in the present work
are due to coalescence of soft partons from the quark-gluon plasma.

\subsection{kaon transverse momentum spectrum and kaon to pion ratio}

\begin{figure}[ht]
\includegraphics[height=3in,width=2.5in,angle=-90]{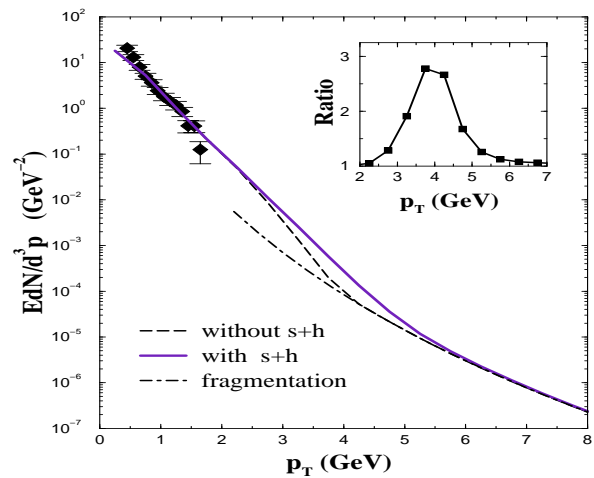}
\vspace{-0.2cm}
\caption{Transverse momentum spectra of $K^-$ from Au+Au collisions
at $\sqrt s=200$ AGeV. The dashed curve includes contributions 
only from coalescence of partons in the quark-gluon plasma and hadrons 
from independent fragmentations of minijet partons (dash-dotted curve). 
Adding also hadrons from coalescence of minijet partons with partons in 
the quark-gluon plasma gives the solid curve. Ratio of the solid to 
the dashed curve is given in the inset. Experimental $K^-$ data 
\cite{phenix} at 130 AGeV are shown by filled circles.}
\label{kaon}
\end{figure}

In Fig. \ref{kaon}, we show the antikaon spectrum including those
from decays of $K^{*-}$ for Au+Au collisions at $\sqrt s=200$ AGeV. 
Results including antikaon production from coalescence of soft partons
from the quark-gluon plasma are shown by dashed curve while those including
also coalescence between minijet partons with the quark-gluon plasmas
are shown by the solid curve. Contributions from minijet fragmentations 
(dash-dotted curve) are also included. As for antiprotons, we compare 
these predictions with the experimental data shown by filled squares from 
the PHENIX collaboration for Au+Au collisions at $\sqrt s=130$ AGeV 
\cite{phenix}, as there are no published data at $\sqrt s=200$ AGeV. 
For the limited data below 2 GeV/c, the coalescence model reproduces them 
very well without the contribution from coalescence of minijet partons 
with partons from the quark-gluon plasma as the latter becomes important 
when the transverse momentum is above 2.5 GeV/c.  Shown in the inset of Fig.
\ref{kaon} is the ratio of the predictions with and without the
contribution from soft and hard parton coalescence as a function of 
transverse momentum. The ratio reaches a factor of close to 3 
at transverse momentum around 4 GeV/c. The contribution to antikaons 
from coalescence of minijet partons with those from the quark-gluon plasma 
is thus comparable to that for pions. 

\begin{figure}[ht]
\includegraphics[height=3in,width=2.5in,angle=-90]{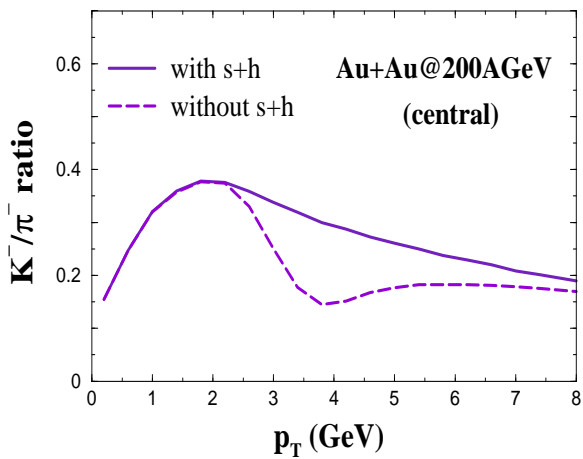}
\vspace{-0.2cm}
\caption{Antikaon to pion ratio from Au+Au collisions at $\sqrt s=200$ 
AGeV. Solid and dashed curves are, respectively, results with and
without contributions to antikaon production from coalescence of minijet 
partons with those from the quark-gluon plasma.} 
\label{ratiok}
\end{figure}

The antikaon to pion ratio is shown in Fig. \ref{ratiok} as a function 
of transverse momentum. The solid curve is the result including 
contributions from both parton coalescence and minijet fragmentations.
For transverse momenta below about 2 GeV/c, this ratio is similar
to the antiproton to pion ratio except that its value is smaller. 
At higher transverse momenta, the antikaon to pion ratio decreases
slightly and reaches the value predicted by pQCD at high transverse
momenta \cite{levai0}, which gives a larger antikaon to pion ratio than
the antiproton to pion ratio. Results without contribution to antikaon 
production from coalescence of minijet partons with partons from the 
quark-gluon plasma are given by the dashed curve, and it gives a smaller 
antikaon to pion ratio at intermediate transverse momenta compared to 
that with this contribution. Our results thus demonstrate again the 
importance of antikaon production from coalescence of partons from 
minijets and quark-gluon plasma.

\section{transverse flow effect}\label{flow}

\begin{figure}[ht]
\includegraphics[height=3in,width=4.0in,angle=-90]{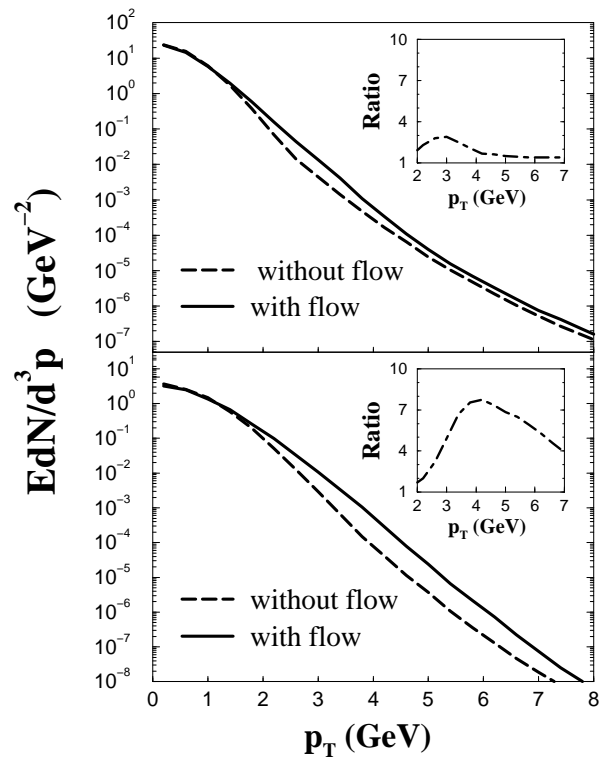}
\vspace{-0.2cm}
\caption{Pion (upper panel) and antiproton (lower panel) transverse momentum 
spectra from coalescence of partons in the quark-gluon plasma with 
(solid curves) and without (dashed curves) collective transverse flow 
in the quark-gluon plasma. Ratio of these two results are shown in the 
insets.}
\label{eff-flow}
\end{figure}

It is interesting to study the effect of transverse flow of quark-gluon 
plasma on the transverse momentum spectra of produced hadrons. In fact,
the increase of antiproton to pion ratio at low transverse momenta was 
believed to be due to a stronger transverse flow effect on protons than 
on pions \cite{heinz}. Such effect was assumed in our previous study 
\cite{greco} by using a larger inverse slope parameter for the antiproton
transverse momentum spectrum than that for pions. In the present model, 
soft hadrons are produced from coalescence of partons in the quark-gluon 
plasma, which is given a collective flow velocity of 0.5c, and the
resulting antiproton to pion ratio at low transverse momentum is found
to increase with transverse momentum. To see if this is due to the
collective flow effect introduced in the model, we have repeated the 
calculations without transverse flow effect on antiprotons.  These results, 
which include only contribution to antiproton production from coalescence 
of partons in the quark-gluon plasma, are shown in Fig. \ref{eff-flow} 
by dashed curves for pions (upper panel) and antiprotons 
(lower panel). Compared with results with a collective flow velocity of 
0.5c, shown by solid curves in Fig. \ref{eff-flow}, collective flow affects 
the pion and antiproton spectra mainly at transverse momenta above 1.5 GeV/c, 
and its effect is stronger for antiprotons than for pions. The reason
that for low transverse momenta the inverse slope parameter for pions 
remains smaller than that for antiprotons in the absence of collective
flow is due to the fact that most low transverse momentum pions are
from decays of rho mesons, which gives a smaller effective slope parameter
for pions than that of directly produced pions. Our results thus 
demonstrate that the increase seen in the antiproton to pion ratio at low
transverse momenta is not necessarily due to the collective flow
of quark-gluon plasma.
 
\begin{figure}[ht]
\includegraphics[height=3in,width=2.5in,angle=-90]{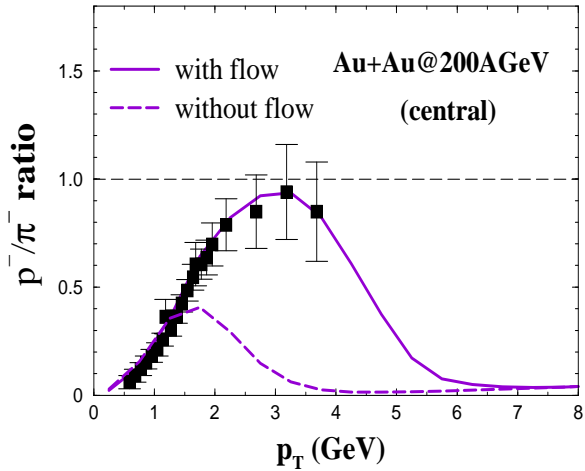}
\vspace{-0.2cm}
\caption{Antiproton to pion ratio with (solid curve) and without (dashed
curve) collective flow in the quark-gluon plasma. Experimental data
are shown by filled squares.}
\label{rationf}
\end{figure}

Since collective flow affects hadron spectra at intermediate transverse
momenta as shown in Fig.\ref{eff-flow} and its insets, where the ratio 
of pion or antiproton spectrum obtained with and without collective flow 
in the quark-gluon plasma is shown, it is expected to have a large
effect on the antiproton to pion ratio at these momenta. This is shown  
in Fig.\ref{rationf}, where the antiproton to pion ratio is given for
both with (solid curve) and without (dashed curve) collective flow
effect on protons. It is seen that collective flow enhances significantly 
this ratio for transverse momenta between 2 and 5 GeV/c. The collective flow
effect on intermediate transverse momentum antiprotons is thus as strong as 
the effect due to coalescence of partons from minijets with those from the 
quark-gluon plasma, shown in Fig. \ref{ratio}. To confirm the mechanism for
antiproton production from coalescence of minijet and quark-gluon plasma
partons, it is thus important to have a quantitative understanding of 
collective flow in the quark-gluon plasma. 

\section{elliptic flows}\label{elliptic}

The parton coalescence model based on the test particle Monte-Carlo
method is also useful for studying other observables at RHIC. 
One problem that can be addressed with this model is to study how elliptic
flows of hadrons are related to that of partons \cite{monlar,lin1}.
We carry out such a study following the idea of Ref. \cite{hwa},
where the parton transverse momentum distribution is extracted from
fitting the pion transverse momentum spectrum using the quark recombination
or coalescence model and is then used predict that of antiprotons.
Here, we extract the elliptic flows of light and strange quarks by fitting 
the measured pion and kaon elliptic flows based on the coalescence model. 
The resulting quark elliptic flows are then used to predict the
elliptic flows of protons, $\Lambda$, $\Omega$, and phi mesons.

The elliptic flow is a measure of the anisotropy of particle transverse
momentum spectrum, i.e., 
\begin{eqnarray}
v_2=\left\langle\frac{p_x^2-p_y^2}{p_x^2+p_y^2}\right\rangle,
\end{eqnarray}
where the transverse axes $x$ and $y$ are, respectively, in and out of 
the reaction plane. Including transverse momentum anisotropy, the quark 
transverse momentum distribution is given by
\begin{eqnarray}
\frac{dN_q}{d^2{\bf p}_{\rm T}}&=&\frac{dN_q}{p_{\rm T}dp_{\rm T}d\phi}
=\frac{dN_q}{p_{\rm T}dp_{\rm T}}\left[1+v_2\cos(2\phi)\right],
\end{eqnarray}
where $\phi$ is the azimuthal angle in the transverse plane.

\begin{figure}[ht]
\includegraphics[height=3in,width=2.5in,angle=-90]{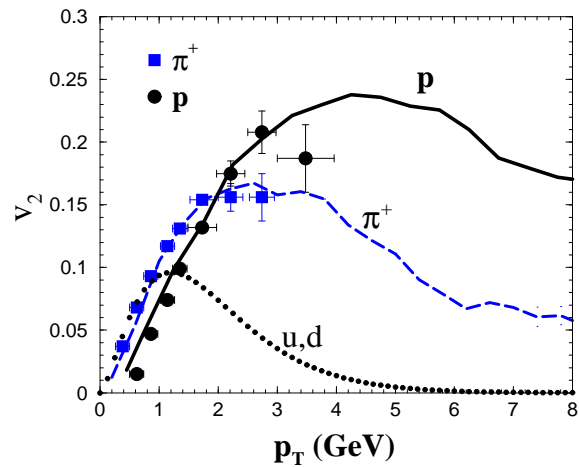}
\vspace{-0.2cm}
\caption{Elliptic flows of pions (dashed curve) and protons (solid curve)
as functions of transverse momentum.  Elliptic flow of light quarks and 
antiquarks is given by dotted curve. Experimental data \cite{esumi} are 
shown by filled squares for pions and filled circles for protons.}
\label{v2p}
\end{figure}

In Fig. \ref{v2p}, we show by dotted curve the elliptic flows of light 
quarks and antiquarks together with the pion elliptic flow shown by 
dashed curve, which is supposed to reproduce the measured pion elliptic
flow given by filled squares \cite{esumi}. The predicted proton elliptic flow 
obtained from the quark and antiquark elliptic flows is then given by the 
dashed curve and is seen to agree with that measured in experiments shown by
filled circles \cite{esumi}. 

\begin{figure}[ht]
\includegraphics[height=3in,width=2.5in,angle=-90]{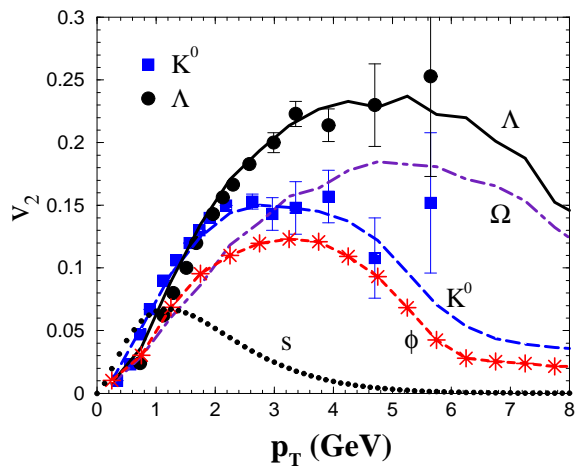}
\vspace{-0.2cm}
\caption{Elliptic flows of kaons (solid curve), phi mesons (dashed
curve with stars), $\Lambda$ (dashed curve), and $\Omega$ (dash-dotted curve). 
Elliptic flow of strange quarks and antiquarks is given by dotted curve. 
Experimental data \cite{snellings} are shown by filled squares for kaons 
and circles for $\Lambda$.}
\label{v2s}
\end{figure}

The elliptic flow of strange hadrons are shown in Fig. \ref{v2s}.
The strange quark and antiquark elliptic flow (dotted curve) used
to fit the measured kaon elliptic flow \cite{snellings} shown by filled 
squares for data and dashed  curve from the coalescence model. The predicted 
$\Lambda$ elliptic flow shown by dashed curve is seen to agree with the 
available experimental data \cite{snellings} given by filled circles. 
We have also predicted the phi meson (dashed curve with stars) and $\Omega$ 
elliptic flows (dash-dotted curve), which are smaller than the kaon and
$\Lambda$ elliptic flow, respectively, as a result of smaller (about 30\%)
strange quark elliptic flows than that of light quarks.

For all hadron considered here, decrease of their elliptic flows at high 
transverse momenta is due to our assumption that high transverse 
momentum minijet partons and hadrons from minijet fragmentations 
have vanishing elliptic flows. Of course, if high transverse momentum
minijet partons also develop elliptic flow as suggested in Ref. 
\cite{gyulassy}, our results on hadron elliptic flows at high transverse 
momenta should be modified.

\section{Summary and outlook}\label{summary}

In this paper, we have studied the hadronization of quark-gluon plasma
and minijet partons produced in relativistic heavy ion collisions 
in terms of the parton coalescence model. The momentum spectra
of partons in the quark-gluon plasma is taken to have an exponential 
form with inverse slope parameter similar to the phase transition
temperature, while partons in the minijets have power-law spectra.
The flavor compositions in the quark-gluon plasma is determined by
the experimentally measured antiproton to proton ratio and 
strange to nonstrange particle ratios. The volume of quark-gluon plasma 
is then fixed by the total transverse energy measured in experiments. 
A collective flow is introduced in the quark-gluon plasma with a flow
velocity comparable to that extracted from experiments. To take into account
the vast difference in the magnitude of the minijet parton transverse
momentum spectrum and that of partons in the quark-gluon plasma, a test
particle Monte-Carlo method has been introduced to efficiently evaluate
the coalescence probability of partons. Both soft partons
from the quark-gluon plasma and hard partons from minijets are used
in the coalescence model to produce hadrons such as pions, kaons (antikaons), 
rho mesons, $K^*$'s, nucleons (antinucleons), and $\Delta$ ($\bar\Delta$)
resonances. Specifically, we have included coalescence of hard minijet 
partons with soft partons besides that among soft and hard partons.

The resulting pion, antikaon, and antiproton spectra are seen to agree 
with available experimental data from RHIC. For pions, contributions from 
rho decays are important in explaining the measured transverse 
momentum spectrum below 2 GeV/c. For intermediate transverse momentum 
spectra between 2 and 5 GeV/c, including contributions from coalescence 
of minijet partons with those from the quark-gluon plasma are important, 
leading to a factor of more than two enhancement compared to results 
without this contribution. We have also compared the transverse momentum 
dependence of antiproton to pion ratio to the experimental data. Results 
from the coalescence model are found to agree quite well with the available
data, i.e., it increases with transverse momentum and reaches a value
of about one at transverse momentum of about 3 GeV/c. With further 
increase in transverse momentum, our model predicts that the antiproton 
to pion ratio should decrease and approach the small value given by 
the pQCD. We have also calculated the antikaon to pion ratio as a function 
of transverse momentum. The result is similar to that for antiproton 
to pion ratio but with a smaller magnitude and a larger value at high 
transverse momenta. We further find that these ratios are reduced if 
antiproton production from coalescence of minijet partons with partons 
in the quark-gluon plasma is neglected. This effect is, however, comparable 
to that due to collective flow of quark-gluon plasma. To confirm the 
mechanism for antiproton production from coalescence of minijet partons 
with partons in the quark-gluon plasma thus requires a quantitative 
understanding of collective flow effects in the quark-gluon plasma. 

We have also studied elliptic flows of hadrons using quark
and antiquark elliptic flows fitted to the measured pion and kaon 
elliptic flows. Predicted proton and $\Lambda$ elliptic flows  
agree with those measured in experiments. We have further
predicted the $\Omega$ elliptic flow, which is smaller
than other hadrons due to smaller strange quark elliptic flow than
that of light quarks.

The parton coalescence model based on the test particle 
Monte-Carlo method can be further extended to include collision dynamics
of partons and hadrons using parton and hadron transport models. 
This would make it possible to study effects due to final-state hadronic 
scattering on the transverse momentum spectra and elliptic flows of hadrons,
which have been neglected in present study as we have compared 
hadron spectra elliptic flows from parton coalescence directly with 
experimental data. Also, including parton scatterings would allow us to treat 
properly the parton spectrum around the cutoff momentum $p_0$, resulting
in a smooth spectrum at this momentum. It further makes it possible
to determine both transverse and elliptic flows of partons from the 
collision dynamics instead of treating them as input as in the present 
study. Moreover, including expansion dynamics of the system will
ensure that the total entropy does not decrease even though 
the entropy density is reduced when hadrons are formed from 
coalescence of partons.

\begin{acknowledgments}
This paper is based on work supported by the U.S. National Science 
Foundation under Grant No. PHY-0098805 and the Welch Foundation under 
Grant No. A-1358. V.G. is also supported by the National Institute of 
Nuclear Physics (INFN) in Italy, while P.L. by the Hungarian OTKA Grant 
Nos. T034269 and T043455. P.L. further thanks G. Papp and G. Fai for 
discussions on pQCD results.
\end{acknowledgments}

\end{document}